\documentclass[%
 reprint,
 amsmath,amssymb,
 aps,
]{revtex4-2}

\usepackage{graphicx}% Include figure files
\usepackage{dcolumn}% Align table columns on decimal point
\usepackage{bm}% bold math
\usepackage{xcolor}
\usepackage{amsmath}

\begin{document}

% User defined commands:
\newcommand\needref{{${}^{\it\bf ref\quad}$}}
\newcommand{\ignoreText}[1]{}
\newcommand{\abssq}[1]{\ensuremath{\left\vert #1\right\vert^2}}
\def\LLP{\psi_{{\textstyle\mathstrut}\tiny L}}
\def\LUP{\chi_{{\textstyle\mathstrut}\tiny L}}
\def\RUP{\chi_{{\textstyle\mathstrut}\tiny R}}
\def\RLP{\psi_{{\textstyle\mathstrut}\tiny R}}
\preprint{APS/123-QED}
\title{Electrical Control of Polariton Josephson Junctions via Exciton Stark Effect}
%\title{Electrical Control of Exciton-Polariton Condensate Josephson Junctions}% Force line breaks with \\
%\thanks{A footnote to the article title}%

\author{Hua Wang}
\email{hua.wang.phys@ou.edu}
\author{Hong-Yi Xie}
\email{hongyi.xie-1@ou.edu}
\author{Kieran Mullen}
\email{kieran@ou.edu}
\affiliation{Homer L. Dodge Department of Physics and Astronomy, Center for Quantum Research and Technology, The University of Oklahoma, Norman, Oklahoma 73069, USA
}%

\date{\today}% It is always \today, today,
             %  but any date may be explicitly specified

\begin{abstract}
We propose harnessing the tools of modern nano-fabrication to
provide electrical control of exciton-polariton (EP) condensates. 
We develop the theory of a device based on the Josephson effect in which electric fields can be used to both switch between and monitor various dynamical modes. 
In particular, both the bias potential and the Josephson energy can be tuned electrically via the exciton component. We model the device by a Gross-Pitaevskii equation assuming that ideal EP condensates are established with well-balanced pumping and dissipation. 
%All the model parameters are calculated microscopically. In particular, we obtain the polariton tunneling amplitude across the junction as a second-order process of electron-hole pair tunneling. 
We find that the EP condensates can be manipulated through degrees of freedom not easily accessible in other coherent quantum systems, and
the dynamics of EP Josephson junctions are richer than that of the conventional superconducting junctions. The ability to control and monitor the condensate by both optical and electrical means allows new ways to study its physics not possible by either, alone.

\end{abstract}

%\keywords{Suggested keywords}%Use showkeys class option if keyword
                              %display desired
\maketitle

%\tableofcontents
%\ignoreText{Exciton-Polaritons in microcavities and Josephson effects in general. Derive from Gross-Pitaevskii equations. Describe the geometry and physical properties of the polariton junctions.}

\emph{Introduction.} 
Quantum coherence has been observed in a host of many-body systems over a broad range of temperatures, from Bose-Einstein condensates of alkali metal vapors at nanokelvins~\cite{WiemanCornell1995,Ketterle1995} to exotic states in cuprates~\cite{Zaanen2015} that persist up to $\sim 130 \, \mathrm{K}$~\cite{Eigler2021}. 
Studies of these systems have enriched our understanding of quantum coherence and correlations~\cite{ChuangNielson}, and stimulated the development of quantum computation~\cite{Martinis2019},  communication~\cite{Kimble2008},  sensing~\cite{Degen2017}, and  simulation~\cite{Gross2017}. 
At the heart of various quantum technologies is the Josephson junction, a device harnessing supercurrents induced by phase differences in macroscopic coherent states of quantum particles. 
A conventional Josephson junction consists of two superconductors separated by a thin insulating layer, where the condensate oscillations driven by a bias voltage are typically a small fraction of the total condensate population due to the high energy cost of a large charge imbalance~\cite{TinkhamMichael1996ItS2}. 
In an analogous system, the bosonic Josephson junction consisting of two weakly coupled Bose-Einstein condensates of ultracold atoms, rich phenomena such as quantum self-trapping effect have been observed due to enhanced particle interactions~\cite{PhysRevLett.79.4950,PhysRevA.59.620,PhysRevLett.95.010402,PhysRevLett.133.093401}. 
Remarkable progress in manipulating these ultracold-atom systems has been achieved based on dynamical light shaping techniques~\cite{Ryu_2015, Rubinsztein-Dunlop_2017, barredo2018}.

Another candidate system for bosonic Josephson junctions is the microcavity exciton-polariton (EP), a hybrid boson that is a coherent superposition of one cavity photon and one exciton, which in turn is a composite boson consisting of a bound electron and hole~\cite{haug2020quantum}. 
%\ignoreText{Their hybridization produces an upper-polariton (UP) and lower-polariton (LP) excitation branch [Fig.~\ref{fig1:sketch}(a)].}
Cavity polaritons exhibit condensation and superfluidity even near room temperature~\cite{Yamamoto2007,Skolnick2010,Amo2014,Krizhanovskii2018,Klembt2018,Xiong2020}, because they have extremely low effective masses and are insensitive to crystal disorder in solids~\cite{kasprzak_boseeinstein_2006,RevModPhys.82.1489,RevModPhys.85.299}. \ignoreText{and thermal equilibrium can be archived in EP condensates}
%\ignoreText{The exciton component of EP is more polarizable than the tightly bound electrons in a ground state atom, and the cavity is intrinsically compatible with nanoelectronics fabrication, making electrical control possible.}
The pioneering experimental studies on the EP Josephson effect are based on trapping the cavity modes~\cite{Schneider_2017}, for example, exploiting the natural double traps due to thickness variations of the cavity~\cite{PhysRevLett.105.120403} or fabricating a double-micropillar structure~\cite{nphysics_pillar_2013}. 
In these systems, the initial state is prepared by large condensation imbalance~\cite{PhysRevLett.105.120403} or by resonate laser pumping~\cite{nphysics_pillar_2013}. The EP Josephson energy is given by the inter-trap tunneling energy of the photonic modes, which is preset via the cavity material and geometry. Moreover, the chemical potential bias induced by non-resonantly pumped excitons is time-independent and indirectly determined by the dynamical balance of short-lived excitons~\cite{nphysics_pillar_2013}. 
Manipulating polaritons via the exciton component~\cite{10.1063/1.2164431,PhysRevB.84.245318,doi:10.1126/science.1140990} allows an alternative, more flexible device design for non-resonantly pumped EP condensates. In condensates dynamically maintained in thermal equilibrium~\cite{PhysRevLett.118.016602,doi:10.1126/sciadv.adk6960}, it is feasible to control the chemical potential bias via the single-exciton potential energy. In particular, \ignoreText{by applying electric fields to semiconductor quantum wells, }the exciton potential energy can be tuned locally via the quantum confined Stark effect~\cite{PhysRevB.76.085304,Schneider_2017}. 

%To some extent the highly nonequilibrium nature of these EP condensates obstructs the application of EP junctions in quantum technology. Thankfully, recent experiments have shown that it is possible to have true thermal equilibrium in EP condensates~\cite{PhysRevLett.118.016602,doi:10.1126/sciadv.adk6960}. 
%\ignoreText{A suitable technique has been developed in ~\cite{10.1063/1.2164431} and successfully applied to trap an exciton-polariton condensate~\cite{doi:10.1126/science.1140990}, where the stress induces a deformation potential in the semiconductor, causing a significant shift in exciton energy while having negligible effect on the photon energy.

\begin{figure}
     \centering
     \includegraphics[width=3.3truein]{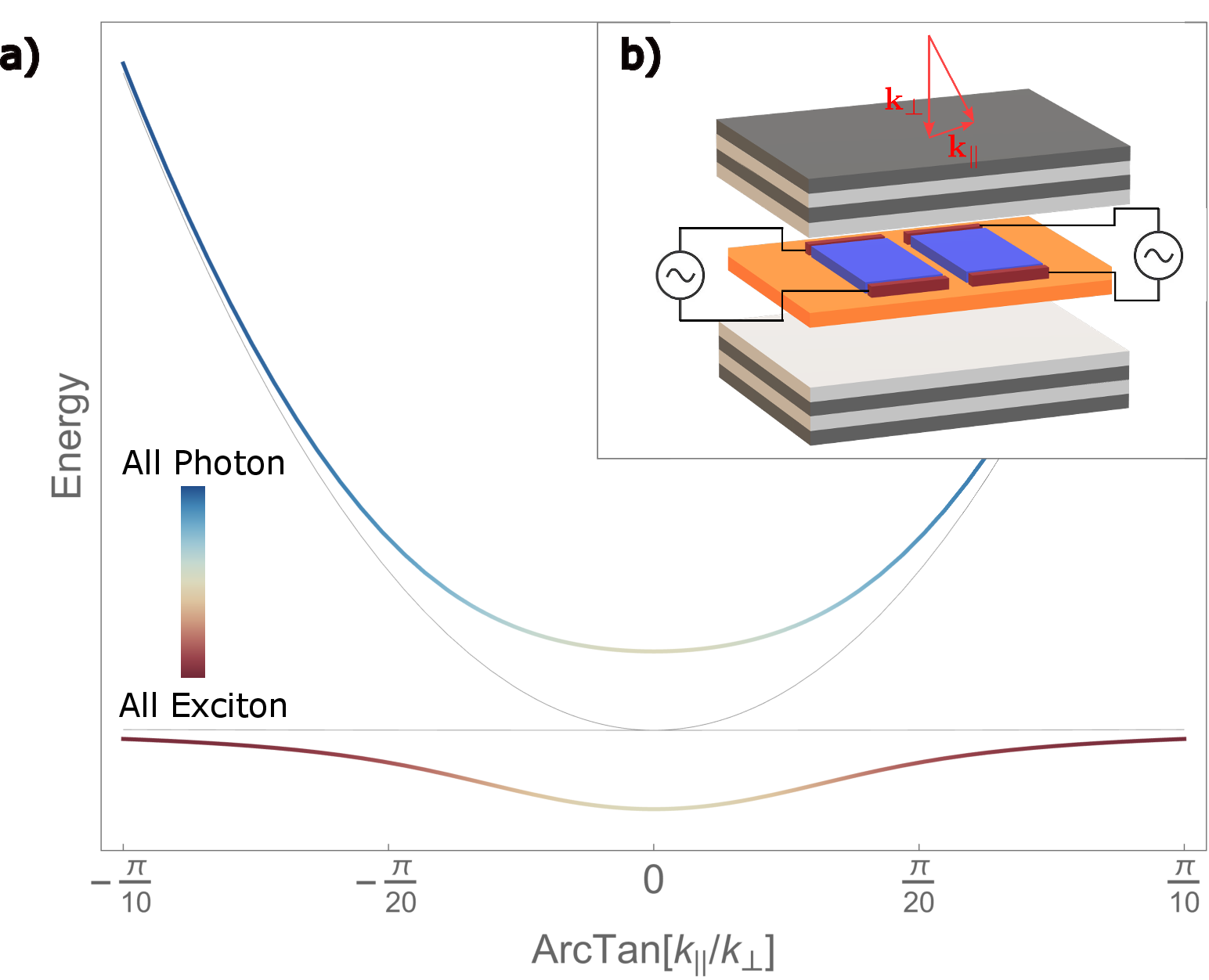}
     \caption{Polariton dispersion relation and schematics of EP Josephson junctions. (a) Typical energy dispersions of cavity polariton along in-plane momentum of incident light $k_{\parallel}$. Upper and lower polariton branches are formed due to strong exciton-photon hybridization at $k_{\parallel} =0$. The color codes of the lower (upper) branch represent the weight of exciton (photon). (b) Schematic geometry of an EP junction. The two-dimensional device is defined via nanofabrication of a semiconducting material embedded in an optical microcavity. %formed by distributed Bragg reflectors and continuously 
     \ignoreText{pumped by a laser} Two electrodes bracket each semiconductor to apply an in-plane electric field, which polarizes the EP condensate as well as to measure the local capacitance (Color online).}
     \label{fig1:sketch}%
\end{figure}

In this letter, we propose a EP Josephson junction device in which the potential energy bias and the Josephson energy can be dynamically controlled by electrical fields.
In the semiconducting layer of the optical microcavity, we introduce a double-well potential to the excitons, which defines two polariton trapping areas. Each area is bracketed between two conductor plates~(see Fig.~\ref{fig1:sketch}), which generate an in-plane electric field controlled by an external voltage source. 
Electrical control in the EP junction allows us to explore rich dynamics not present in the conventional superconducting junction or the photonic double-well trap. We find that it is possible to switch between different macroscopic quantum states, to tune the system from periodic to chaotic behavior. The added electrical elements also provide new methods to probe the system that complement the measurements based on optical coherence~\cite{PhysRevLett.99.126403,PhysRevB.81.033307}.

%\ignoreText{ As shown in Fig.~\ref{fig1:sketch}(b), we consider two semiconductor thin layers horizontally separated by a tunneling barrier. Each layer is encapsulated between two conductor plates and an in-plane electric field can be generated and controlled by an external voltage source. The semiconductor device is embedded in an optical microcavity and one longitudinal cavity mode within the semiconductor band gap, which can hybridize with the subgap excitons.}

%A quantum well is positioned between the two Bragg mirrors of an optical microcavity, where the semiconductor layer hosts two-dimensional excitons. 

%\ignoreText{The condensate current through this area arises from the correlated tunneling of an electron and hole, analogous to the original Josephson effect based on the correlated tunneling of Cooper pairs. Experimental evidence for this mechanism has been found in a few exciton-based systems~\cite{wang_evidence_2019,doan_signature_2023}.}

%, and \textcolor{purple}{that the system can sustain multiple types of condensates(we admit the upper polariton is lossy, can we still sustain it?)}.  \ignoreText{Having {\it both} electrical and optical probes enables control and characterization not possible with optical techniques alone.} 

\emph{Four-component Gross-Pitaevskii equation.}
We start by expanding the  cavity photon state around its longitudinal mode ($\mathbf{k_\perp}$ perpendicular to the layers), resulting in a parabolic dispersion along $\mathbf{k}_{\parallel}$ with a small effective mass $m \sim 10^{-5}m_e$~\cite{RevModPhys.82.1489}, where $m_e$ is the mass of a free electron. It strongly couples to the quasi-2D states of excitons in the semiconducting layer, forming the lower polaritons (LP) and upper polaritons (UP). To emphasize the effect of both branches, we formulate and plot the zero detuning case [see Fig.~\ref{fig1:sketch}(a)], where both LP and UP are equal-weighted superpositions of a photon and an exciton near $\mathbf{k}_{\parallel}=0$ (A general theory can be found in the Supplementary Material). The EP condensation of either branch occurs when the states in the vicinity of $\mathbf{k_{\parallel}}=0$ are macroscopically occupied. The time scales for the condensate formation and decay are discussed at the end of the paper.

We assume that the exciton trapping areas are sufficiently small so that the spatial variations of the condensate density within either one can be neglected. The UP and LP condensates confined on the left and right side of the device can then be described by a four-component Gross-Pitaevskii equation (GPE). The detailed derivation is presented in Supplementary Material. We introduce the complex variables $\psi_\kappa$ and $\chi_\kappa$ to describe the macroscopic amplitude of the polariton ground state $\phi_{\kappa}(\mathbf{r})$ for the LP and UP condensates, respectively, where $\kappa \in \{L,R\}$ denotes the left (L) and right (R) side of the junction. The time evolution of the lower polariton condensates is described by
\begin{eqnarray}
    i \hbar \frac{\partial \LLP}{\partial t} &=&  
    \left( -\frac{\mu_{{\textstyle\mathstrut}LR}+\mu_{{\textstyle\mathstrut}UL}}{2} + \frac{U}{4} \left\vert \LLP \right\vert^2 +
   \frac{U}{2} \left\vert \LUP \right\vert^2\right) \LLP 
   \nonumber \\ 
    &+& \bigg\{ U    \left[\frac{1}{4} \LLP^*  \LUP +  \frac{1}{2} \left(\left\vert  \LLP \right\vert^2
      +  \left\vert  \LUP \right\vert^2\right) \right]  -I_1\bigg\}\LUP
      \nonumber \\
      &-& J \RLP +I_2 \RUP , \nonumber \\
       i \hbar \frac{\partial \RLP}{\partial t} &=&  
    \left( \frac{\mu_{{\textstyle\mathstrut}LR}-\mu_{{\textstyle\mathstrut}UL}}{2} +\frac{U}{4} \left\vert  \RLP \right\vert^2 +
   \frac{U}{2} \left\vert  \RUP \right\vert^2\right)  \RLP 
   \nonumber \\
    &+&  \bigg\{ U \left[\frac{1}{4}  \RLP^*  \RUP +  \frac{1}{2} \left(\left\vert  \RLP \right\vert^2
      +  \left\vert  \RUP \right\vert^2\right) \right]-I_1\bigg\}  \RUP
      \nonumber \\
      &-& J \LLP +I_2 \LUP ,  \label{eq:condEqn}
\end{eqnarray}
where $\mu_{{\textstyle\mathstrut} LR}$ is the energy difference between the left and right LP condensates, $\mu_{{\textstyle\mathstrut}UL}$ is the energy difference between LP and UP condensates on one side, $J$ is the Josephson energy, and $U$ is the interaction energy arising from the scattering of the exciton components of polaritons. Two similar equations describe the upper EP condensate, obtained by exchanging $\psi \leftrightarrow \chi $ and replacing $\mu_{{\textstyle\mathstrut}UL} \rightarrow -\mu_{{\textstyle\mathstrut}UL}$. 

The energy difference $\mu_{{\textstyle\mathstrut}UL}$ is determined by the vacuum Rabi splitting at $\mathbf{k}_\parallel=0$. 
The left-right bias potential $\mu_{{\textstyle\mathstrut} LR}$ is controlled by the in-plane electric fields applied to the two semiconductors via the exciton Stark effect. The exciton components are polarizable, and the polarizability $\alpha$ can be estimated by perturbation theory in terms of exciton orbital wavefunctions, which resembles that of a 2D hydrogen atom~\cite{PhysRevA.43.1186}
$$
\alpha=-e^2\sum_{n,l=\pm 1}\frac{|\langle \Phi_{10}|\hat{\mathbf{x}}|\Phi_{nl} \rangle|^2}{E_{10}-E_{nl}}
$$
where $\Phi_{nl}$ is the 2D electron-hole orbital wavefunction in principle quantum number $n$ and angular quantum number $l$, and $E_{nl}<0$ is the corresponding eigen-energy, $\hat{\mathbf{x}}$ is the e-h displacement operator in the direction of the applied electric field ${\cal E}_{L/R}$. This produces the energy difference  $\mu_{LR}= \alpha\left( {\cal E}_L^2-{\cal E}_R^2\right)/2$. For excitons in GaAs~\cite{chuang2009physics}, we estimate $\alpha \sim 10^{-3} e^2 \mathrm{\mu m^2 eV^{-1}}$. Including the dielectric constant of the host semiconductor, an electric field of  $\sim 1\, V/\mu m$ is needed to create a dipole energy of $\sim$ 0.1 meV.

%\ignoreText{The polariton tunneling energy $J$ is purely determined by exciton tunneling processes through the inverse Hopfield transformation~\cite{RevModPhys.82.1489}. Up to second order in single-electron/hole tunneling strength, we obtain
%\begin{align}
%    J= & \, -t_e t_h \sum_\mathbf{k} \Phi^\ast_{\nu',\mathbf{K}}(\mathbf{k})\Phi_{\nu,\mathbf{K}}(\mathbf{k}) \nonumber \\
%   & \times \left( \frac{1}{E_{\nu \mathbf{K}}-\epsilon_{\mathbf{k},\mathbf{K}-\mathbf{k}}}+\frac{1}{E_{\nu'\mathbf{K}}-\epsilon_{\mathbf{k},\mathbf{K}-\mathbf{k}}}\right). \label{epj}
%\end{align}
%Here $\Phi_{\nu,\mathbf{K}}(\mathbf{k})$ is the Fourier component of the wavefunction of the 2D exciton in $\nu$-orbital, with $\mathbf{K}$ being the center-of-mass momentum, $E_{\nu \mathbf{K}}$ is the exciton-band energy, $\epsilon_{\mathbf{k},\mathbf{k}'}$ is the energy of an electron at momentum $\mathbf{k}$ and hole at momentum $\mathbf{k}'$ in the absence of interactions, and $t_{e,h}$ are the single-particle tunneling energy for the electrons and holes (see Supplemental Material). For 2D excitons in 1s state,
%we evaluate $J = (8 \pi^2/3) (t_e t_h)/(|E_{10}|)$, where $|E_{10}|=\hbar^2/(2\mu a^2)$ is the binding energy, with $a$ being the 2D Bohr radius and $\mu$ the reduced mass of an electron-hole pair. In principle, the Josephson energy in our proposed device can be engineered to a wide range as it depends on both electron and hole tunneling, and remains highly tunable if the barrier potential is further controlled by gate voltage.}

The Josephson energy $J$ arises from the spatial overlapping of left and right trapped polariton states. Assuming that the excitons are confined in a double square well potential, we estimate the Josephson energy as a function of the exciton tunneling barrier 
\begin{equation} \label{je}
J \approx \pi^2 \frac{D_b}{D_w} E_0 e^{-D_b/\xi},    
\end{equation}
where $D_w$ and $D_b$ are the widths of a single well and the tunneling barrier, respectively, $E_0 = \frac{\hbar^2 \pi^2}{2 m D_w^2}$ with $m$ being the polariton effective mass, and $\xi =\hbar/\sqrt{m V_B}$ is the penetration depth with $V_B$ being the tunneling barrier strength. The calculation is shown in Supplementary Material. We note that Eq.~\eqref{je} is valid in the tunneling regime $D_b \gg \xi$. 
The Josephson energy tunability $| J(V_B+\delta V)/J(V_B)-1| \sim \frac{D_b}{\xi} \frac{\delta V}{V_B}$ for $\delta V \ll V_B$, which can be of order of one due to the large prefactor $D_b/\xi$. In addition, the coupling energies between upper and lower polaritons $I_{1,2}$ are determined by the spatial overlapping of the localized LP and UP wavefunctions, which can be neglected for strong Rabi splitting $\mu_{LR} \gg I_{1,2}$.

The interaction energy $U$ arises from the $s$-wave scattering of the excitons. For the double-square-well potential we estimate $U \approx 3 g_s /(2 D_w)$, where $g_s = 6 |E_{10}| a_\mathrm{B}^2$ is the contact interaction parameter with $a_\mathrm{B}$ being the Bohr radius and $E_{10}$ the $1s$ state energy~\cite{PhysRevB.59.10830}. The magnitude of $g_s$ is usually characterized by the blue shift of LP dispersion~\cite{sun_direct_2017}. In a GaAs-based 2D layer with unspecified geometry we estimate $g_s$ of the order of $1 \sim 10 \, \mu \mathrm{eV} \mu \mathrm{m}^2$. The tunability of the Josephson energy $J$ in Eq.~\eqref{je} can make it possible to study the interplay between the Josephson effect and the non-linearity induced by interactions.

\emph{Results.} We consider two major cases. In the first, the coherent excitation in the upper polariton branch is negligible, and the coupled equations reduce to the two-component bosonic junction model that has been intensively studied~\cite{PhysRevA.59.620}. Defining the dimensionless interaction strength $\Lambda=U N_T/2J$ and time $\tau=\frac{2J}{\hbar}t$ in LP parameters, where $N_T=|\psi_L|^2+|\psi_R|^2$, and parametrizing
$\psi_{L,R}=\sqrt{N_T(1\pm z)/2}e^{i\theta_{L,R}}$ with $|z| \le 1$, we transform Eq.~\eqref{eq:condEqn} to
\begin{equation} \label{2-comp-GP}
    \dot{\theta}=\Lambda\Big(z-z_b(\tau)\Big)+\frac{z\cos\theta}{\sqrt{1-z^2}}, \quad \dot{z}=-\sqrt{1-z^2}\sin \theta,
\end{equation}
where $\theta = \theta_R-\theta_L$. 
%We note that the conjugate variables $\{z,\theta\}$ form the phase space of the dynamical system. 

\begin{figure}
    \centering
    \includegraphics[width=3.3truein]{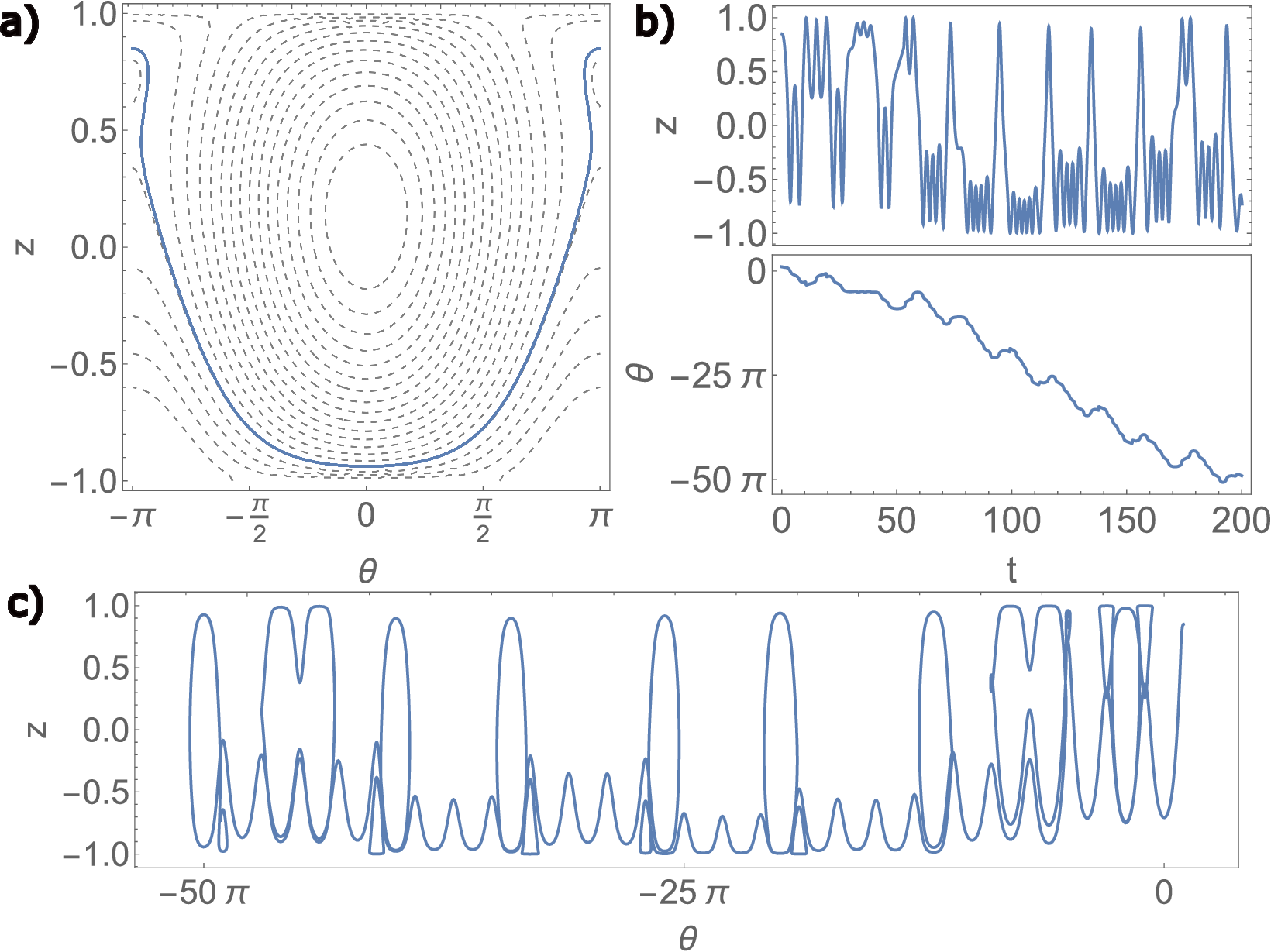}
    \caption{Stable and chaotic dynamics in LP-only Josephson junctions [Eq.~\eqref{2-comp-GP}]. We take $\Lambda=2$ and $z_0=0.2$. (a) Energy contour in phase space for static driving potential $z_1=0$. For the initial conditions $z(0)=0.85$ and $\theta(0)=\pi$, the system follows the blue trajectory. (b) Time evolution of density (upper panel) and phase (lower panel) for dynamical driving potential $z_1 = 0.4$. The phase-space trajectory exhibits chaotic behavior as shown in (c).}
    \label{fig2:2com}
\end{figure}

For a time-independent bias potential $z_b(\tau)=z_0$, such a system resembles an undriven, non-linear pendulum. Its dynamics fall into two categories: those for which the trajectory is a closed loop in phase space; and those in which a pendulum would swing through the complete circle, so the trajectory is unbounded. The latter are termed ``macroscopic quantum self-trapping'' modes~\cite{PhysRevA.59.620}, meaning the condensate density difference remains non-zero. Both types of modes are present in the energy contours [Fig.~\ref{fig2:2com}(a)]. When the bias potential has a constant and an oscillating component $z_b(\tau)= z_0 + z_1 \sin(\Omega \tau)$, it is possible to switch between different trajectories within the same dynamical category. 
However, the modes near the separatrix are unstable, and the dynamical potential can lead to chaotic behavior [Fig.~\ref{fig2:2com}(b)].
 
The second major case is that both the LP and UP condensates are non-negligible so that the system must be described by the four-component model. Switching to the dimensionless variables
$g=U/J$, $\delta=\mu_{LR}/J$, $\Delta=\mu_{UL}/J$, and $\lambda_{1,2}=I_{1,2}/J$, and performing a rotating-frame transformation (see Supplemental Material), 
we rewrite Eq.~\eqref{eq:condEqn} in the matrix form
\begin{equation} \label{4-comp-block}
    i\partial_\tau
    \begin{pmatrix}        
        \tilde{\psi}_L\\
        \tilde{\psi}_R\\
        \tilde{\chi}_L\\
        \tilde{\chi}_R\\
    \end{pmatrix}=
    \begin{pmatrix}
        h_{LP} &v\\
        v^\dagger &h_{UP}
    \end{pmatrix}
    \begin{pmatrix}        
        \tilde{\psi}_L\\
        \tilde{\psi}_R\\
        \tilde{\chi}_L\\
        \tilde{\chi}_R\\
    \end{pmatrix}
\end{equation}
where the blocks of the effective Hamiltonian read
\begin{equation}
\begin{aligned}
& h_{LP} =
\begin{pmatrix}
    \frac{\delta}{2}+\frac{g}{4}|\tilde{\psi}_L|^2+\frac{g}{2}|\tilde{\chi}_L|^2 &-1\\
    -1 &-\frac{\delta}{2}+\frac{g}{4}|\tilde{\psi}_R|^2+\frac{g}{2}|\tilde{\chi}_R|^2\\
\end{pmatrix},\nonumber \\
& h_{UP} =
\begin{pmatrix}
    \frac{\delta}{2}+\frac{g}{4}|\tilde{\chi}_L|^2+\frac{g}{2}|\tilde{\psi}_L|^2 &-1\\
    -1 &-\frac{\delta}{2}+\frac{g}{4}|\tilde{\chi}_R|^2+\frac{g}{2}|\tilde{\psi}_R|^2\\
\end{pmatrix},\nonumber \\
& v= e^{-i\Delta \tau}
\begin{pmatrix}
    \frac{g}{2}n_{L}+\lambda_1 & \lambda_2\\
    \lambda_2        &\frac{g}{2}n_R+\lambda_1
\end{pmatrix} +e^{-i2\Delta \tau}
\begin{pmatrix}
    \frac{g}{4}\tilde{\psi}_L^*\tilde{\chi}_L &0\\
    0   &\frac{g}{4}\tilde{\psi}_R^*\tilde{\chi}_R\\
\end{pmatrix},\\
\end{aligned}
\end{equation}
where $n_{L(R)}$ is the total particle number on the left (right) side.

In an EP system, the LP-UP energy difference $\mu_{UL}$, which is the Rabi-splitting determined by strong photon-exciton hybridization, is several orders of magnitude larger than the left-right energy difference $\mu_{LR}$. In Eq.~\eqref{4-comp-block}, for $\Delta \gg \delta,g,\lambda_{1,2}$, the inter-branch block $v$ carries a fast oscillating phase. Under the random phase approximation, we neglect this inter-branch coupling and obtain a pair of two-component equations for upper and lower EP condensates,
\begin{equation} \label{4-comp-sep}
 i\partial_\tau
    \begin{pmatrix}
        \tilde{\psi}_L\\
        \tilde{\psi}_R\\
    \end{pmatrix}= h_L \begin{pmatrix}
        \tilde{\psi}_L\\
        \tilde{\psi}_R\\
    \end{pmatrix},  \quad i\partial_\tau
    \begin{pmatrix}
        \tilde{\chi}_L\\
        \tilde{\chi}_R\\
    \end{pmatrix}= h_U \begin{pmatrix}
        \tilde{\chi}_L\\
        \tilde{\chi}_R\\
    \end{pmatrix}.
\end{equation} 
We note that the polariton-exchange process between LP and UP branches is effectively absent, and the particle numbers of UP and LP branches are approximately conserved in the time scale much larger than Rabi period. However, the existence of the upper polariton serves as a dynamical chemical potential to the lower polariton condensate, causing the density $n_{L/U}$ in lower/upper polariton only oscillate at a small amplitude around its mean value.
For the LP condensates, we parametrize $\psi_{L(R)}=\sqrt{n_{LP}(1\pm z_{LP})/2}e^{i\theta_{LP,L(R)}}$ where $n_{LP}$ is the total number of lower polaritons and $|z_{LP}|\le 1$; Similar parametrization applies to the UP condensates. We transform Eq.~\eqref{4-comp-sep} to
\begin{align}
& \dot{\theta}_\sigma=\Lambda_\sigma(z_\sigma-z_{b,\sigma})+\frac{z_\sigma}{\sqrt{1-z_\sigma^2}}\cos\theta_\sigma, \nonumber \\
& \dot{z}_\sigma=-\sqrt{1-z_b^2}\sin \theta_\sigma,
\end{align}
where $\sigma \in \{LP,UP\}$, $\theta_\sigma = \theta_{\sigma,L}-\theta_{\sigma,R}$, and the dynamical chemical potential bias $z_{b,\sigma}=z_{0,\sigma}+c_\sigma z_{\bar{\sigma}}$ consists of two parts:  $z_{0,\sigma}=-\frac{\delta}{2\Lambda_\sigma}$ with $\Lambda_\sigma=g n_\sigma/4$  and $c_\sigma=n_{\bar{\sigma}}/n_\sigma$, the over-bar refers to the opposite EP component.  The chemical potential in one branch is dynamically influenced by its counterpart.

\begin{figure}
    \centering
    \includegraphics[width=3.5truein]{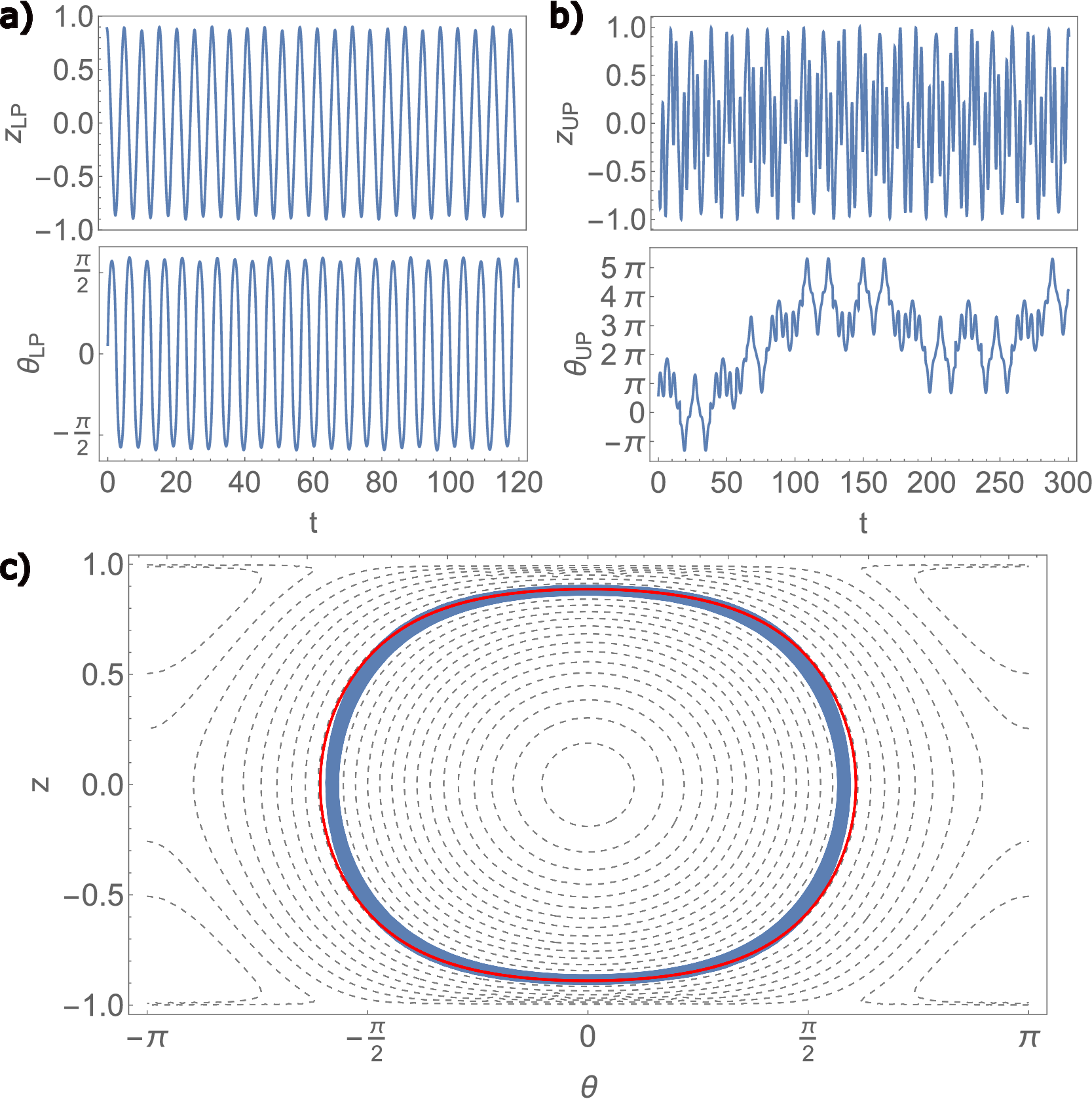}
    \caption{Dynamics of LP and UP condensates for low UP filling. We take $n_{LP}=0.95$ and $n_{UP}=0.05$. (a) and (b) Time evolution of LP and UP condensates, respectively. (c) Phase-space trajectory of LP condensates corresponding to (a). The red trace is a closed trajectory for stable state of two-component LP-only model  (Color online).}
    \label{fig3:periodic}
\end{figure}

\begin{figure}
    \centering
    \includegraphics[width=3.3truein]{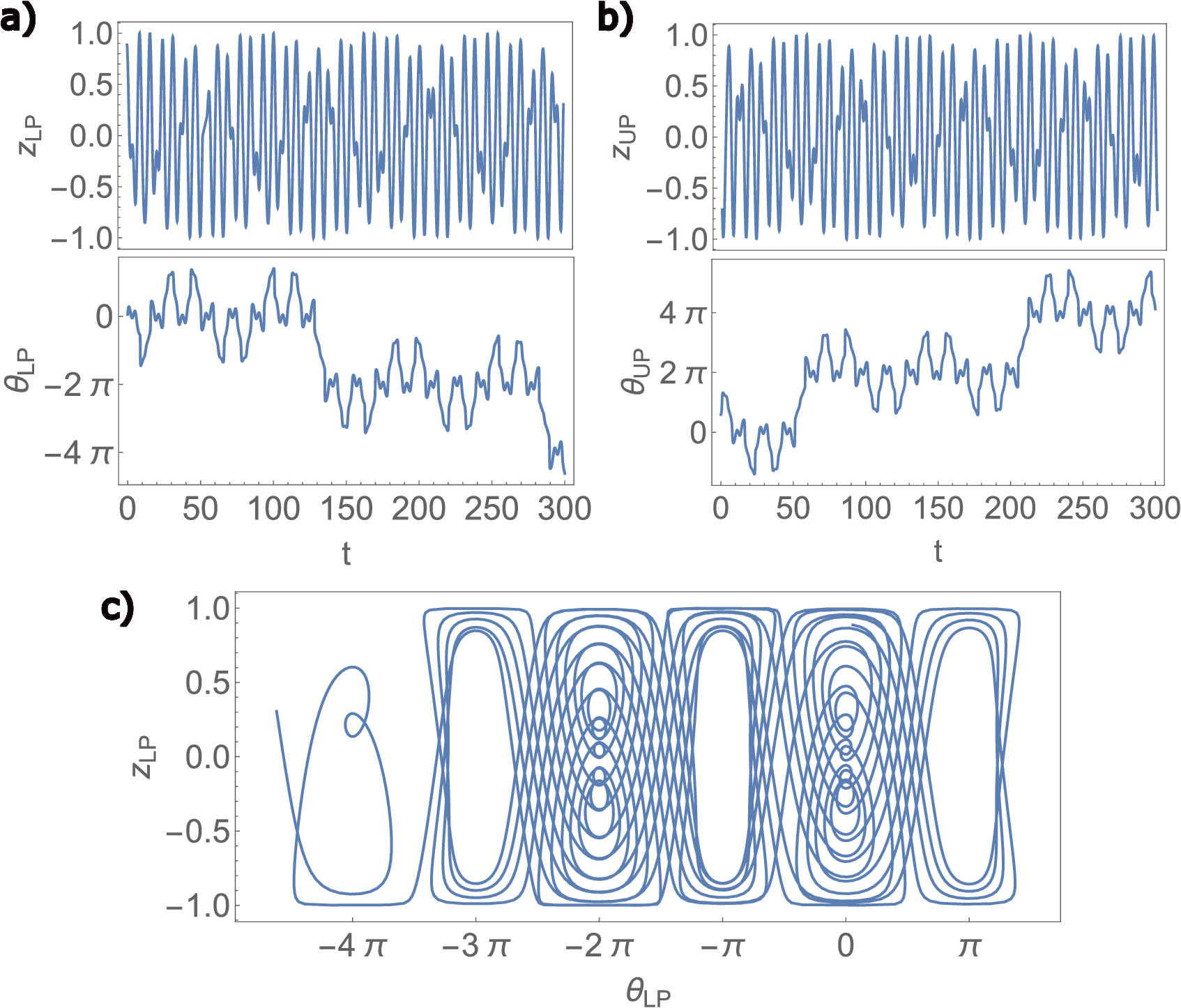}
    \caption{Dynamics of LP and UP condensates for equal UP and LP fillings $n_{LP}=n_{UP}=0.5$. (a) and (b) Time evolution of LP and UP condensates, respectively. (c) Phase-space trajectory of LP condensates corresponding to (a).}
    \label{fig4:chaos}
\end{figure}

When the UP branches are insufficiently populated $n_{LP} \gg n_{UP}$, as shown in Fig.~\ref{fig3:periodic}, we find that the LP condensates exhibit high-frequency but small-amplitude oscillations that are induced by the UP condensates. However, the UP condensates exhibit small-amplitude but chaotic oscillations due to the periodic driving force generated by the LP condensates through interactions. We justify that the two-component GP equation~\eqref{2-comp-GP} is valid only for LP condensates in this limit.

In contrast, when the population of the two EP condensates are comparable, the temporal oscillations in one component can act as a dynamical driving force in the other. We always obtain chaotic behavior for both LP and UP condensates, as shown in Fig.~\ref{fig4:chaos}.

\emph{Discussion.} 
In this letter we have explored the wealth of possible behavior when nanofabricated electrical elements are incorporated into EP junctions, which provide additional and simultaneous probes of the condensates. Nanofabrication has enabled the creation of single electron devices, the monitoring of single spins, and the control of superconducting qubits. We believe that its incorporation into optical cavities will lead to a better understanding of EP condensates.

Our EP junction model~\eqref{eq:condEqn} assumes the existence of well-defined and conserved condensates in steady states, since we have neglected the imaginary source and sink terms describing the coupling between the condensates and a population reservoir~\cite{PhysRevLett.99.140402,RevModPhys.85.299}, and its validity depends on time scales. 
If the re-population and decay rates are slow compared to the Josephson frequency, 
the gain and loss processes are balanced and all the dynamical behaviors predicted in our work should be observed. 
Otherwise, if the decay rate is much larger than the Josephson frequency, the condensates are unstable and the oscillations are damped~\cite{PhysRevLett.99.140402,PhysRevLett.105.120403,nphysics_pillar_2013}; 
if the re-population is much larger than the Josephson frequency, the oscillations are absent and a steady and phase-dependent flow of polaritons occurs across the junction. 
 
In a non-equilibrium polariton gas, the chemical potential bias of the EP junction depends on the dynamics of the population reservoir, which is difficult to determine or control directly. Thankfully, recent experiments have demonstrated that true thermal equilibrium can develop in lower polaritons with a long lifetime $\sim 270 \, \mathrm{ps}$~\cite{PhysRevLett.118.016602,doi:10.1126/sciadv.adk6960}, and directly controlling the chemical potential via the Stark effect is possible for lower polaritons. The upper polaritons are usually very lossy, and the UP condensate can be formed through resonant pumping~\cite{doi:10.1021/acs.nanolett.3c03123}. Nevertheless, as we have shown, the LP condensate is not influenced by the UP branches as long as the density of the pumped UP condensate is much lower than that of the spontaneously formed LP condensate. 

The EP junction model~\eqref{eq:condEqn} also neglects the incoherent tunneling of uncondensed thermal polaritons due to finite temperature. 
This incoherent tunneling gives rise to a dissipative current in addition to the supercurrent of EP condensates, in analogy to the quasi-particle tunneling in superconducting Josephson junctions. We develop the corresponding ``resistively shunted junction model'' elsewhere~\cite{TBP}.

We have not considered the pseudo-spin of polaritons in this work. 
It has been predicted that for LP condensates the spin degrees of freedom allow a chaotic behavior in the Josephson oscillations~\cite{PhysRevB.80.235303}.
We expect that spin relaxation~\cite{PhysRevB.47.15776} could also couple the EP condensates to dark exciton states~\cite{PhysRevLett.99.176403}. 
This mechanism may introduce even richer dynamics and provide a potential technique for detecting dark excitons via their influence on the capacitance. Small variations in capacitance can be measured very precisely
by their effect on the resonance frequency of co-fabricated
circuit elements. 
In the self-trapping phase, there is a constant shift in the capacitance, and, in the oscillating phase, the capacitance shift would oscillate at the Josephson frequency. 
The capacitance can provide information on the total condensate density, including the possibility of observing ``dark'' condensates that are not optically observable. 

We thank B.~Uchoa, A.~Auerbach, Y.~Zhang, H.~Alnatah, A.~Javadi, and Y.~Li for stimulating discussions. The work of H.-Y.X. is supported by the Dodge Family Fellowship granted by the University of Oklahoma. 

\appendix

\end{document}

\title{SUPPLEMENTAL MATERIAL: \\
Electrical Control of Polariton Josephson Junctions via Exciton Stark Effect}

\author{Hua Wang}
\email{hua.wang.phys@ou.edu}
\author{Hong-Yi Xie}
 \email{hongyi.xie-1@ou.edu}
 \author{Kieran Mullen}
\email{kieran@ou.edu}
\affiliation{Homer L. Dodge Department of Physics and Astronomy, Center for Quantum Research and Technology, The University of Oklahoma, Norman, Oklahoma 73069, USA}

%%%%%%
\maketitle
%%%%%%
\onecolumngrid
%\pagestyle{empty}

%%%%%%
\section{Josephson energy and polariton-polariton interactions}
\subsection{General theory}
We consider a polariton system where the exciton component is subject to a confinement potential. In momentum space, the exciton-photon Hamiltonian reads
\begin{equation} \label{ham-0}
    H_0= \sum_{\mathbf{k},\mathbf{k}'} 
   \begin{pmatrix}
   a_\mathbf{k}^\dagger & b_\mathbf{k}^\dagger    
    \end{pmatrix} \begin{pmatrix}
    \frac{\hbar^2k^2}{2m_\mathrm{ex}}\delta_{\mathbf{k},\mathbf{k'}} +V_\mathrm{ex}(\mathbf{k}-\mathbf{k'}) & \Omega \, \delta_{\mathbf{k},\mathbf{k'}} \\
    \Omega \, \delta_{\mathbf{k},\mathbf{k'}} & \left(\frac{\hbar^2k^2}{2m_{\mathrm{ph}}} + \Delta \right) \delta_{\mathbf{k},\mathbf{k'}}
    \end{pmatrix}
    \begin{pmatrix}
   a_{\mathbf{k}'} \\ b_{\mathbf{k}'}   
    \end{pmatrix}, 
\end{equation}
where $a_\mathbf{k}^\dagger$ and $b_\mathbf{k}^\dagger$ ($a_\mathbf{k}$ and $b_\mathbf{k}$) are the exciton and cavity photon creation (annihilation) operators, respectively, $m_\mathrm{ex}$ and $m_\mathrm{ph}$ are the exciton and photon masses, respectively, $\Delta=E_\mathrm{ph}-E_\mathrm{ex}$ is the detuning energy, $V_\mathrm{ex}(\mathbf{k})$ is the Fourier transform of the exciton confinement potential $V_\mathrm{ex}(\mathbf{r})$, and  $\Omega = \Phi_{10}(0) \, |\mathbf{d}_{cv}| \sqrt{E_\mathrm{ph}/2\varepsilon_0\varepsilon_r}$ is the exciton-photon coupling energy, with $\Phi_{10}(\mathbf{\bfr})$ being the envelope function of an exciton in $1s$ orbit, and $\mathbf{d}_{cv}$ the electric dipole moment of an electron-hole excitation in the semiconductor~\cite{haug2020quantum}. Polariton interactions originate from the exciton component. We assume that the exciton interaction Hamiltonian takes the form
\begin{equation} \label{hint}
    H_\text{int}= \frac{g_s}{\mathcal{A}} \sum_{\bk,\bk',\bq} a_\bk^\dagger a_{\bk'}^\dagger a_{\bk'+\bq} a_{\bk-\bq},
\end{equation}
where $g_s = 6|E_{10}|a_\mathrm{B}^2$ is the contact interaction parameter, with $|E_{10}|$ being the $1s$ state binding energy and $a_\mathrm{B}$ the Bohr radius, and $\mathcal{A}$ are the thickness and the total area of the quantum well, respectively~\cite{PhysRevB.59.10830}.     

In the absence of the confinement potential $V_\mathrm{ex} =0$, the Hamiltonian~\eqref{ham-0} can be diagonalized by the Hopfield transformation
\begin{equation}
    \hat{O}_\mathbf{k}=
    \begin{pmatrix}
        X_\mathbf{k} & C_\mathbf{k}\\
        -C_\mathbf{k} & X_\mathbf{k}\\
    \end{pmatrix}, \quad
    X_\mathbf{k}=\sqrt{\frac{1}{2}\bigg(1+\frac{\Delta_\mathbf{k}}{\sqrt{\Delta_\mathbf{k}^2+4\Omega^2}}\bigg)}, \quad C_\mathbf{k}=\sqrt{\frac{1}{2}\bigg(1-\frac{\Delta_\mathbf{k}}{\sqrt{\Delta_\mathbf{k}^2+4\Omega^2}}\bigg)},
\end{equation}
where $\Delta_\mathbf{k}=\D + \frac{\hbar^2k^2}{2}\Big(\frac{1}{m_\mathrm{ph}}-\frac{1}{m_\mathrm{ex}}\Big)$ is the direct gap between photon and exciton modes. The polariton modes are defined via the transformation
\be \label{pl-b}
\begin{pmatrix}
   P_{\mathbf{k}} \\ Q_{\mathbf{k}} \end{pmatrix} = \hat{O}_\mathbf{k}^T
   \begin{pmatrix}
   a_{\mathbf{k}} \\ b_{\mathbf{k}}    
    \end{pmatrix},
\ee
and its Hermitian conjugation, where $P_\bk$ and $Q_\bk$ denote the annihilation operators of the lower (LP) and upper (UP) polaritons, respectively. In polariton basis \eqref{pl-b}, the Hamiltonian \eqref{ham-0} is written as
\begin{equation} \label{ham-1}
    H_0= \sum_{\mathbf{k},\mathbf{k}'} 
   \begin{pmatrix}
   P_\mathbf{k}^\dagger & Q_\mathbf{k}^\dagger    
    \end{pmatrix} \left[ \begin{pmatrix}
    \vep_{\mathrm{LP},\bk} & 0 \\
    0 & \vep_{\mathrm{UP},\bk}
    \end{pmatrix} \delta_{\mathbf{k},\mathbf{k'}} + 
    \begin{pmatrix}
    X_{\bk} X_{\bk'} & X_{\bk} C_{\bk'} \\
    C_{\bk} X_{\bk'} & C_{\bk} C_{\bk'} 
    \end{pmatrix} V_\mathrm{ex}(\mathbf{k}-\mathbf{k'}) \right]
    \begin{pmatrix} 
   P_{\mathbf{k}'} \\ Q_{\mathbf{k}'}   
    \end{pmatrix},
\end{equation} 
where the polariton dispersions read 
\be \label{pl-d} 
\vep_{\mathrm{LP(UP)},\bk} = \frac{\vep_\bk -(+) \sqrt{\Delta_\mathbf{k}^2+4\Omega^2}}{2}, \quad \vep_\bk = \D + \frac{\hbar^2k^2}{2}\Big(\frac{1}{m_\mathrm{ph}}+\frac{1}{m_\mathrm{ex}}\Big).
\ee

\emph{Effective polariton Hamiltonian.} We note that in Eq.~\eqref{ham-1} the confinement potential $\propto V_\mathrm{ex}$ is nonlocal in coordinate space because it involves Hopfield coefficients at different momenta. However, since we are interested in the EP condensates about $\bk=0$, for strong exciton-photon coupling $\W \gg |\D_{k_c}-\D|$, where $k_c$ is the typical width of the condensate momentum distribution, we can approximate 
$X_{\bk} \approx X_0$ and $C_{\bk} \approx C_0$. Moreover, the polariton dispersions \eqref{pl-d} can be approximated by the quadratic forms $\vep_{\mathrm{LP(UP)},\bk} \approx \ep_{\mathrm{LP(UP)}} + \frac{\hbar^2 k^2}{ 2 m_{\mathrm{LP}(UP)}}$, where the zero-point energies $\ep_{\mathrm{LP(UP)}} = \frac{\D -(+)\sqrt{\D^2 + 4 \W^2}}{2}$ and the inverse effective masses
\be
\frac{1}{m_\mathrm{LP}}=\frac{X_0^2}{m_\mathrm{ex}}+\frac{C_0^2}{m_\mathrm{ph}}, \quad     \frac{1}{m_\mathrm{UP}}=\frac{C_0^2}{m_\mathrm{ex}}+\frac{X_0^2}{m_\mathrm{ph}}.
\ee
Under this long-wavelength approximation, we obtain the effective polariton Hamiltonian in coordinate space:
\begin{align} \label{ham-2}
& H_0= \int d^2\bfr 
   \begin{pmatrix}
   P_\bfr^\dagger & Q_\bfr^\dagger    
    \end{pmatrix} 
    \begin{pmatrix}
        h_{\mathrm{LP}}(\bfr)  & \td{V}(\mathbf{r})\\
        \td{V}(\mathbf{r}) & h_{\mathrm{UP}}(\bfr) 
    \end{pmatrix}
    \begin{pmatrix} 
   P_\bfr \\ Q_\bfr  
    \end{pmatrix}, \quad  h_{\zeta}(\bfr) =
        -\frac{\hbar^2 \nabla^2}{2m_\mathrm{\zeta}} + \ep_{\mathrm{\zeta}} + V_\mathrm{\zeta}(\mathbf{r}),
\end{align}
where $\zeta \in \{\mathrm{LP,UP}\}$, $V_\mathrm{LP}(\bfr) = X_0^2 V_\mathrm{ex}(\bfr)$, 
$V_\mathrm{UP}(\bfr) = C_0^2 V_\mathrm{ex}(\bfr)$, and
$\td{V}(\bfr) = X_0C_0 V_\mathrm{ex}(\bfr)$.

\subsection{Four-state approximation}

We assume that the exciton confinement potential $V_\mathrm{ex}(\bfr)$ is a double-well trap. For large barrier potential, the left (right) well can be approximated by a single well $V_{\mathrm{ex},L(R)}(\bfr)$. For $V_{\mathrm{ex},\k}(\bfr)$ with $\k \in \{L,R\}$, one can solve the ground state wave function $\td{\phi}_{\zeta,\k}(\mathbf{r})$ of a $\zeta$ polariton (a real function). We note that $\td{\phi}_{\zeta,L}(\mathbf{r})$ and $\td{\phi}_{\zeta,R}(\mathbf{r})$ are almost orthogonal for a large potential barrier due to exponentially small overlaps. 
Via the Gram–Schmidt process one can obtain a pair of localized basis wavefunctions $\phi_{\zeta,\k}(\mathbf{r}) \approx \td{\phi}_{\zeta,\k}(\mathbf{r})$ with $\int \! d^2 \bfr\,  \phi_{\zeta,\k}(\mathbf{r})  \phi_{\zeta,\k'}(\mathbf{r}) = \del_{\k,\k'}$. We project the polariton states onto the localized basis
\be \label{pj}
P_\bfr \approx \sum_{\k = L,R} \phi_{\mathrm{LP},\k}(\mathbf{r}) \, P_{\k}, \quad 
Q_\bfr \approx \sum_{\k = L,R} \phi_{\mathrm{UP},\k}(\mathbf{r}) \, Q_{\k}.
\ee
and the Hermitian conjugations, where $P_{\k}$ and $Q_{\k}$ are polariton annihilation operators in the localized states. The effective polariton Hamiltonian \eqref{ham-2} takes the four-component form 
\be \label{ham-3}
H_0 = 
\begin{pmatrix}
   P_L^\dagger & P_R^\dagger & Q_L^\dagger & Q_R^\dagger    
\end{pmatrix} 
\begin{pmatrix} 
\vep_{\mathrm{LP},L} & J_{\mathrm{LP}} & I_{LL} & I_{LR} \\
J_{\mathrm{LP}}      &  \vep_{\mathrm{LP},R} & I_{RL} & I_{RR} \\
I_{LL} & I_{RL}  & \vep_{\mathrm{UP},L} & J_{\mathrm{UP}} \\
I_{LR} & I_{RR} & J_{\mathrm{UP}}      &  \vep_{\mathrm{UP},R} \\
\end{pmatrix}
\begin{pmatrix}
   P_L \\ P_R \\ Q_L \\ Q_R   
\end{pmatrix},
\ee
where the matrix elements are given by
\be
\vep_{\zeta,\k} = \int \! d^2 \bfr\, \phi_{\zeta,\k}(\mathbf{r}) \, h_{\zeta}(\bfr) \, \phi_{\zeta,\k}(\mathbf{r}), \quad 
J_{\zeta} = -\int \! d^2 \bfr\, \phi_{\zeta,L}(\mathbf{r}) \, h_{\zeta}(\bfr) \, \phi_{\zeta,R}(\mathbf{r}), \quad
I_{\k\k'} = \int \! d^2 \bfr\, \phi_{\mathrm{LP}, \k}(\mathbf{r}) \, \td{V}(\bfr) \, \phi_{\mathrm{UP},\k'}(\mathbf{r}).
\ee
For large Rabi splitting, $\W \gg |I_{\k\k'}|$, we can neglect the couplings between lower and upper polaritons $I_{\k\k'} \approx 0$.

Via the Hopfield transformation~\eqref{pl-b} and the projection \eqref{pj}, we obtain, 
\be \label{ak-pq}
a_\bk= \sum_{\k \in L,R} \left( X_0 \phi_{\mathrm{LP},\k}(\bk) \, P_\k + C_0\phi_{\mathrm{UP},\k}(\bk) \, Q_\k \right),
\ee
and its Hermitian conjugation, where $\phi_{\zeta,\k}(\bk)$ is the Fourier transform of $\phi_{\zeta,\k}(\bfr)$. Via Eq.~\eqref{ak-pq} the interaction Hamiltonian \eqref{hint} reduces to 
\begin{align}
    H_\text{int}= &\,  \frac{g_s}{l_z} \sum_{\k} \big[ 
     c_{\k,0} \, \hat{P}_\k^\dagger \hat{P}_\k^\dagger \hat{P}_\k \hat{P}_\k
  + 2c_{\k,1} \, ( \hat{P}_\k^\dagger \hat{P}_\k^\dagger \hat{P}_\k \hat{Q}_\k + \hat{P}_\k^\dagger \hat{Q}_\k^\dagger \hat{P}_\k \hat{P}_\k)
  + c_{\k,2} \, (\hat{P}_\k^\dagger \hat{P}_\k^\dagger \hat{Q}_\k \hat{Q}_\k + \hat{Q}_\k^\dagger \hat{Q}_\k^\dagger \hat{P}_\k \hat{P}_\k + 4 \hat{P}_\k^\dagger \hat{Q}_\k^\dagger \hat{P}_\k \hat{Q}_\k ) \nonumber \\
    &\,  + 2c_{\k,3} \, (\hat{P}_\k^\dagger \hat{Q}_\k^\dagger \hat{Q}_\k \hat{Q}_\k + \hat{Q}_\k^\dagger \hat{Q}_\k^\dagger \hat{P}_\k \hat{Q}_\k)
    + c_{\k,4} \, \hat{Q}_\k^\dagger \hat{Q}_\k^\dagger \hat{Q}_\k \hat{Q}_\k
    \big], \label{hint-1}
\end{align}
where $c_{\k,n} \equiv  X_0^{4-n} C_{0}^n \int d^2\bfr\, \phi_{\mathrm{LP},\k}^{4-n}(\bfr)  \phi_{\mathrm{LP},\k}^{n}(\bfr)$. Here we have neglected the inter-well interactions because of the locality $\phi_{\zeta,L}(\bfr) \phi_{\zeta',R}(\bfr) \approx 0$. 

\subsection{Double-square-well confinement potential}
For zero-detuning $\D=0$, we have $X_0=C_0=1/\sqrt{2}$, $V_\mathrm{LP,UP}(\bfr)=\td{V}(\bfr)=V_\mathrm{ex}(\bfr)/2$, and $m_\mathrm{LP}=m_\mathrm{UP} \equiv m$. The lower and upper polariton Hamiltonians in Eq.~\eqref{ham-2} are identical up to the Rabi splitting, $h_{\mathrm{LP(UP)}}(\bfr) =
       h(\bfr) -(+) \, \W $, where
       \be
       h(\bfr) = -\frac{\hbar^2 \nabla^2}{2m} + \frac{1}{2} V_\mathrm{ex}(\mathbf{r}).
       \ee
We model the exciton confinement potential by an asymmetric double square wells
\be
V_{\mathrm{ex}}(x,y) = \begin{cases} V_{L}(x), \quad & x < 0 \\ V_{R}(x), \quad & x >0 \end{cases}, \quad 
\ee
where
\be \label{VRL}
V_{L}(x) = 
\begin{cases} 
0, \quad & x < -D_w-\frac{D_b}{2}, \\
-\mathcal{V}_L, \quad & -D_w-\frac{D_b}{2} < x < -\frac{D_b}{2}, \\
\mathcal{V}_B, \quad & x > -\frac{D_b}{2},
\end{cases} \quad
V_{R}(x) = 
\begin{cases} 
0, \quad & x > D_w+\frac{D_b}{2}, \\
-\mathcal{V}_R, \quad &  \frac{D_b}{2} < x < D_w + \frac{D_b}{2}, \\
\mathcal{V}_B, \quad &  x < \frac{D_b}{2},
\end{cases}
\ee
with $\mathcal{V}_{R,L,B} > 0$. We are interested in the strong confinement regime $\xi \ll D$, where $D \equiv \mathrm{min}\{D_w,D_b\}$ and 
$\xi \equiv \mathrm{max}\{\xi_{L,1},\xi_{L,2},\xi_{R,1},\xi_{R,2}\}$ with $\xi_{\k,1} = \frac{\hbar}{\sqrt{m \mathcal{V}_\k}}$ and $\xi_{\k,2} = \frac{\hbar}{\sqrt{m (\mathcal{V}_\k+\mathcal{V}_B})}$. For the confinement potential $V_\k(x)$ in Eq.~\eqref{VRL}, up to leading order in $\xi/D$, the ground-state energy $E_\k$ and wavefunction $\phi_\k(x)$ of $h(\bfr)$ take the simple expressions
\begin{align}
& E_L \approx -\frac{\mathcal{V}_L}{2} + E_0, \quad \phi_L(\bfr) \approx \sqrt{\frac{2}{l_y D_w}} \times 
\begin{cases} 
\sin(\del_{L,1}) \, e^{(x+D_w+D_b/2)/\xi_{L,1}}, \quad & x < - D_w -\frac{D_b}{2},  \\
\sin\left[\frac{(\pi-\del_{L,1}-\del_{L,2})(x+D_w+D_b/2)}{D_w}+\del_{L,1}\right], & -D_w-\frac{D_b}{2} < x < -\frac{D_b}{2}, \\
\sin(\del_{L,2}) \, e^{-(x+D_b/2)/\xi_{L,2}}, \quad & x > -\frac{D_b}{2}, \\
\end{cases} \nn \\
& E_R \approx -\frac{\mathcal{V}_R}{2} + E_0, \quad \phi_R(\bfr) 
\approx \sqrt{\frac{2}{l_y D_w}} \times 
\begin{cases} 
\sin(\del_{R,2}) \, e^{(x-D_b/2)/\xi_{R,2}}, \quad & x < \frac{D_b}{2},  \\
\sin\left[\frac{(\pi-\del_{R,1}-\del_{R,2})(x-D_b/2)}{D_w}+\del_{L,2}\right], & \frac{D_b}{2} < x < D_w+\frac{D_b}{2}, \\
\sin(\del_{R,1}) \, e^{-(x-D_w-D_b/2)/\xi_{R,1}}, \quad & x > D_w+\frac{D_b}{2}, \\
\end{cases}
\end{align}
where $E_0 = \frac{\hbar^2 \pi^2}{2 m D_w^2}$, $\del_{\k,1(2)}=\pi \xi_{\k,1(2)}/D_w \ll 1$, and $l_y$ is the typical dimension of the quantum well in $y$ direction. We estimate the matrix elements of the effective Hamiltonian \eqref{ham-3}:
\begin{align}
& \vep_{\mathrm{LP},L} = -\mathcal{V} - \frac{\mu_{LR} + \mu_\mathrm{UL}}{2}, \quad
\vep_{\mathrm{LP},R} = -\mathcal{V} + \frac{\mu_{LR} - \mu_\mathrm{UL}}{2}, \quad
\vep_{\mathrm{UP},L} = -\mathcal{V} - \frac{\mu_{LR} - \mu_\mathrm{UL}}{2}, \quad
\vep_{\mathrm{UP},R} = -\mathcal{V} + \frac{\mu_{LR} + \mu_\mathrm{UL}}{2}, \nn \\
& J \equiv -J_\mathrm{LP,UP} = \left[ \sqrt{(\mathcal{V}+\mathcal{V}_B)^2-\mu_{LR}^2} -  \mathcal{V}_B\right] \frac{\pi^2 \xi_{L,2}\xi_{R,2}}{D_w^2} \frac{e^{-\frac{D_d}{\xi_{L,2}}}-e^{-\frac{D_d}{\xi_{R,2}}}}{\frac{D_w}{\xi_{R,2}}-\frac{D_w}{\xi_{L,2}}} \approx
\pi^2 \mathcal{V} \frac{D_b \xi^2}{D_w^3} e^{-D_b/\xi} + \mathcal{O}(\mu_{LR}^2/\mathcal{V}^2), \nn \\
& I_1 \equiv -I_{\k\k} \approx \mathcal{V}, \quad 
  I_2 \equiv I_{LR,RL} \approx \pi^2 \mathcal{V}_B \frac{D_b \xi^2}{D_w^3} e^{-D_b/\xi} + \mathcal{O}(\mu_{LR}^2/\mathcal{V}^2), \label{paras}
\end{align} 
where we have defined 
\be
\mu_{LR} = \frac{\mathcal{V}_L -\mathcal{V}_R}{4}, \quad 
\mu_\mathrm{UL} = 2 \W, \quad 
\mathcal{V} = \frac{\mathcal{V}_L +\mathcal{V}_R}{4}, \quad 
\xi = \frac{\hbar}{\sqrt{m (2\mathcal{V}+\mathcal{V}_B})}.
\ee

We draw important conclusions from Eq.~\eqref{paras}. (i) For weak inter-well potential imbalance $|\mu_{LR}| \ll \mathcal{V}$, the correction to the tunneling energy $J$ is of second order in $|\mu_{LR}/\mathcal{V}|$, which can be neglected. 
(ii) The tunneling energy is sensitive to the potential barrier $\mathcal{V}_B$. We obtain the tunability $|\frac{J(\mathcal{V}_B)-J(0)}{J(0}| \sim \frac{D_d}{\xi} \frac{\mathcal{V}_B}{\mathcal{V}}$ for $\mathcal{V}_B \ll \mathcal{V}$, which can be of order of one due to the large prefactor $D_d/\xi$. 
(iii) For large Rabi splitting, $\mu_\mathrm{UL} \gg \mathcal{V}_{L,R,B}$, we can neglect the couplings between lower and upper polaritons $I_{\k\k'} \approx 0$. 
In the interaction Hamiltonian \eqref{hint-1}, 
\be
c_{k,n} = \frac{1}{4} \int d^2\bfr\, \phi_{\k}^4 (\bfr) \approx \frac{1}{4}\times\frac{3}{2 D_w}.
\ee
In the EP superfluid phase, taking the mean-field approximation $P_\k \sim \psi_{\k}$, $Q_\k \sim \chi_\k$, $P_\k^\dagger \sim \psi_{\k}^\ast$, and $Q_\k^\dagger \sim \chi_\k^\ast$,
from Eqs.~\eqref{ham-3} and \eqref{hint-1} we obtain the GP equation (1) in the main text, where the effective interaction strength reads
\be 
U = \frac{3g_s}{2D_w}.
\ee

\section{GP equation in rotating frame}
The four-component GP equation reads 
$i \partial_t \Psi = H[\Psi^*,\Psi] \, \Psi$, where $\Psi = \begin{pmatrix} \psi_L & \psi_R & \chi_L & \chi_R \end{pmatrix}^T$ and the effective Hamiltonian takes the block form $H = \begin{pmatrix} H_{LP} & V \\ V^\dagger & H_{UP} \end{pmatrix}$, where
\begin{equation}
\begin{aligned}
	&H_{LP}=
	\begin{pmatrix}
		\frac{\delta-\Delta}{2}+\frac{g}{4}|\psi_L|^2+\frac{g}{2}|\chi_L|^2 &-1\\
		-1 &-\frac{\delta+\Delta}{2}+\frac{g}{4}|\psi_R|^2+\frac{g}{2}|\chi_R|^2\\
		\end{pmatrix}, \quad V=
	\begin{pmatrix}
		\frac{g}{4}\psi_L^*\chi_L+\frac{g}{2}n_L+\lambda_1	& \lambda_2\\
		\lambda_2	&\frac{g}{4}\psi_R^*\chi_R+\frac{g}{2}n_R+\lambda_1\\
	\end{pmatrix}, 	\\
	& H_{UP}=
		\begin{pmatrix}
		\frac{\delta+\Delta}{2}+\frac{g}{4}|\chi_L|^2+\frac{g}{2}|\psi_L|^2 &-1\\
		-1 &-\frac{\delta-\Delta}{2}+\frac{g}{4}|\chi_R|^2+\frac{g}{2}|\psi_R|^2\\
	\end{pmatrix}
\end{aligned}
\end{equation}	
with $n_\kappa=|\psi_\kappa|^2+|\chi_\kappa|^2$ being the total density on $\k \in \{L,R\}$ side. For $\Delta \gg \del,g,\lambda_{1,2}$, we decompose the Hamiltonian into two parts $H=H_{\Delta}+H'$, where 
\begin{equation}
H_{\Delta}= \mathrm{diag}\{
		-\Delta/2, -\Delta/2, \Delta/2, \Delta/2 \}.
\end{equation}
Performing a ``rotating frame'' transformation, $\begin{pmatrix} \td{\psi}_L & \td{\psi}_R & \td{\chi}_L & \td{\chi}_R \end{pmatrix}^T \equiv e^{-iH_{\Delta}t} \Psi$ and $\begin{pmatrix} h_{LP} & v \\ v^\dagger & h_{UP} \end{pmatrix} \equiv e^{i H_\D t}H' e^{-iH_\D t}$, we transform the GP equation to Eq.~(4).

%\bibliography{reference_clean}
%